\documentclass{ifacconf}

\usepackage{graphicx}      % include this line if your document contains figures
\usepackage{natbib}        % required for bibliography
\RequirePackage{etex} 

\usepackage[dvipsnames]{xcolor}

\usepackage{amsmath}
\usepackage{amssymb}

\usepackage{enumitem}

\usepackage[normalem]{ulem}

\usepackage{tikz}
\usetikzlibrary{calc}
\usetikzlibrary{shapes}
\usetikzlibrary{decorations.pathreplacing}

\usepackage{changepage}
\usepackage{psfrag}
\usepackage{graphics} % for pdf, bitmapped graphics files
\usepackage{epsfig} % for postscript graphics files
\usepackage{lpic}
\usepackage{amssymb}
\usepackage{textcomp}
\usetikzlibrary{calc,arrows}
\usepackage{etex}
\usepackage{amsmath}
\usepackage{amssymb}
\usepackage{amsfonts}
\usepackage{siunitx}
\usepackage{subcaption}
\usepackage{arydshln}
\usepackage{url}
\usepackage{tabularx}
\usepackage{caption}
\usepackage{times} % assumes new font selection scheme installed
\usepackage{pgfplots}

\newcommand{\bma}[2]       {\left[\begin{array}{#1}#2\end{array}\right]}
\newcommand{\w}        {q}

\usepackage{amsfonts, amssymb,mathrsfs,dsfont,bbm}

\newcommand{\diag}{\mathrm{diag}}
\newcommand{\tr}{{\mathsf{T}}}
\newcommand{\GP}[2]{{\mathcal{G}\!\mathcal{P}}(#1,#2)}
\newcommand{\vc}[1]{\mathrm{#1}}
\newcommand{\mx}[1]{\mathrm{#1}}

\newcommand{\Dcal}{\mathcal{D}} % Training data set
\newcommand{\Tcal}{\mathcal{\boldsymbol{D}}} % Data set (training and validation) 
\newcommand{\Vcal}{v}           % Validation data set
\newcommand{\Ncal}{\mathcal{N}}
\newcommand{\Xcal}{\mathcal{X}}

\newcommand{\Nbb}{\mathbb{N}}
\newcommand{\Rbb}{\mathbb{R}}
\newcommand{\bbk}{k}

\newcommand{\ki}{\kappa}
\newcommand{\enom}{e^{\mathrm{r,nom}}}
\newcommand{\eident}{e^{\mathrm{r,ident}}}

\newcommand*{\argmaxOp}{\operatornamewithlimits{argmax}\limits}

  \DeclareMathAlphabet{\mathbbold}{U}{bbold}{m}{n}
 
\newcommand{\eye}{\mathbb{I}}

\usepackage{algorithm, algorithmicx}%, multirow, titlecaps
\usepackage[english]{babel}
\usepackage[noend]{algpseudocode}

\newtheorem{remark}{Remark}

\begin{document}
\begin{frontmatter}

\title{Parameter Identification for Digital Fabrication: \\ A Gaussian Process Learning Approach} 

\thanks[footnoteinfo]{This project has received funding from the European Union’s Horizon 2020 research and innovation programme under the Marie Sklodowska-Curie grant agreement No.\ 846421. 
This research project is also part of the Swiss Competence Center for Energy Research SCCER FEEB\&D of the Swiss Innovation Agency Innosuisse. 
}

\author[First]{Yvonne R.\ St\"urz}
\author[Second]{Mohammad Khosravi}
\author[Second]{Roy S.\ Smith}

\address[First]{
	Model Predictive Control Laboratory, University of California, Berkeley, CA 94709, USA (e-mail: y.stuerz@berkeley.edu)}
\address[Second]{
	Automatic Control Laboratory, ETH Z\"urich, ZH 8092, Switzerland (e-mail: \{khosravm, rsmith\}@control.ee.ethz.ch).
}

\begin{abstract}                % Abstract of not more than 250 words. 
:~Tensioned cable nets can be used as supporting structures for the efficient construction of lightweight building elements, such as thin concrete shell structures. To guarantee important mechanical properties of the latter, the tolerances on deviations of the tensioned cable net geometry from the desired target form are very tight. Therefore, the form needs to be readjusted on the construction site. In order to employ model-based optimization techniques, the precise identification of important uncertain model parameters of the cable net system is required. 
This paper proposes the use of Gaussian process regression to learn the function that maps the cable net geometry to the uncertain parameters. In contrast to previously proposed methods, this approach requires only a single form measurement for the identification of the cable net model parameters. This is beneficial since measurements of the cable net form on the construction site are very expensive. For the training of the Gaussian processes, simulated data is efficiently computed via convex programming. The effectiveness of the proposed method and the impact of the precise identification of the parameters on the form of the cable net are demonstrated in numerical experiments on a quarter-scale prototype of a roof structure. 
%Numerical experiments on a quarter-scale prototype of a roof structure are presented,  which demonstrate t
%The effectiveness of the method and the impact of the precise identification of the parameters on the form of the cable net have been demonstrated in numerical experiments. 
%
%The impact of the precise parameter identification on the precision of the model-based form of the cable net, and thus the importance of the identification for model-based control is illustrated. 
%
%, and thus the importance of the parameter identification for model-based control
%
%The impact of the precise parameter identification on the precision of the model-based form of the cable net is shown. 
\end{abstract}

\begin{keyword}
Gaussian Process Regression, Learning, Parameter Identification, Optimization, Digital Fabrication, Flexible Formwork, Lightweight Construction
\end{keyword}

% to do for final submission: 
% - add reference to Mohammad's paper: GP for controller parameter tuning 
% - make Figure 2 for 3D plot prettier

\end{frontmatter}
%===============================================================================

\section{Introduction}
\label{sec:intro}

Doubly curved thin concrete shell structures are very efficient building structures, as they can be designed with a high stiffness and structural stability \citep{Tahmoorian2020, Palagi2020, Yang2019}. 
Comparatively little material is needed to span large areas, e.g.\ for roof structures.  %
The construction process requires a so-called formwork, which is a structure supporting the concrete and defining the form of the shell. Once the concrete has cured, the formwork is removed. 
Traditional formwork is usually fabricated using customized wooden parts and is therefore \smash{time-,} labor- and cost-intensive. It furthermore requires a lot of material and produces a lot of waste. 
An alternative flexible formwork can be used instead, which consists of a steel cable net tensioned inside a rigid frame. The concrete is applied on top of the prestressed cable net and a fabric layer. The amount of material used and waste produced can considerably be reduced w.r.t.\ traditional formwork. 

As the structural properties of the concreted shell crucially depend on its form, the tolerances for the cable net geometry are very tight. In order to meet these tolerances, the form of the pretensioned cable net needs to be readjusted on the construction site. This can be done by changing the lengths of the boundary edges. 
Prior work in \citep{Stuerz2016a} has introduced a model-based feedback control strategy which minimizes the deviations of the cable net geometry from the designed target geometry. 
The form of the cable net in static equilibrium can be mathematically described by nonlinear implicit equations representing force balances at all interior nodes of the cable net. Alternatively, for fixed parameters and boundary conditions, the form in static equilibrium can be computed by convex programming \citep{Lobo1998}, \citep{boyd2004convex} 
as a minimum energy state of the system. 

It is known that some model parameters to which the cable net form is sensitive are subject to high fabrication variations. These parameters are the unstressed lengths of the cable edge segments. 
In order for the control method to perform well, either a robust control approach is required \citep{Stuerz2020}, or the important uncertain parameters need to be precisely identified. 
To do so, in \citep{Stuerz2016b}, methods were proposed, which are based on multiple measurements of the as-built cable net geometry. 
Measurements on the construction site are very time-consuming and expensive and their number should thus be minimized. 
To this end, both the identification step and the control step were combined into an iterative closed-loop algorithm in \citep{Stuerz2019}, where after each form measurement, the parameters are re-identified to update the model which is then used to compute the next control inputs. In this way, measurement information and model knowledge can be exploited. 
In \citep{Stuerz2019, Liew2018, Stuerz2019b}, the proposed control strategy was validated on experiments on a quarter-scale prototype of a doubly curved roof shell.

The contributions of this paper are the following: We present a method for the identification of the important uncertain parameters of unstressed cable edge lengths based on Gaussian process (GP) regression. 
While the function mapping the cable net form to the uncertain parameters is not known analytically, we exploit the fact that the inverse function can be formulated as a convex optimization problem based on a minimum energy approach. 
This allows us to efficiently generate training data in simulation, based on which the mapping from the cable net geometry to the parameters is learned. 
Using the trained GP, only one set of form measurements from the construction site is required to identify the uncertain parameters, rather than performing a complete identification procedure \citep{LjungBooK2}.

The paper is structured as follows. Section~\ref{sec:preliminaries} introduces preliminaries about GP regression. In Section~\ref{sec:model}, the model of the cable net system is presented and the problem formulation is stated. Section~\ref{sec:GPlearning} gives the main result of the paper, before numerical experiments are given in Section~\ref{sec:numerics}. 

\textbf{Notation:} 
The sets of natural numbers and of natural numbers up to $n$ are denoted by $\Nbb$ and $\Nbb_n$, respectively. Similarly, the set of real numbers, the $n$-dimensional Euclidean space, and the set of ${n \times m}$ matrices with real entries are denoted by $\Rbb$, $\Rbb^n$ and $\Rbb^{n\times m}$, respectively. 
The identity matrix is written as $\eye$. 
For any vector $\vc{x}$, the Euclidean norm of $\vc{x}$ is written by $\|\cdot\|$.
The diagonal matrix with diagonal entries equal to those of the vector $\vc{x}$ is denoted by $\diag(\vc{x})$. The uniform distribution with support in the interval $[a,b]$ is denoted by $\mathcal{U}(a,b)$.

\section{Preliminaries on Gaussian Process Regression} 
\label{sec:preliminaries}
In this section, we introduce some background on Gaussian process regression which is of relevance for the paper. 

\subsection{Gaussian Processes}\label{ssec:GP}
Let $\Xcal$ be a given set and $\Rbb^\Xcal$ denote the space of $\Rbb$-valued functions defined over $\Xcal$. 
A GP \citep{rasmussen2006gaussian} is a random object with values in $\Rbb^\Xcal$ with the specific property that its restriction to any finite subset of $\Xcal$ induces a normal distribution on the Euclidean space.
More precisely,  let $\mu$ be a function as ${\mu:\Xcal\to\Rbb}$ and $\bbk:\Xcal\times\Xcal\to \Rbb$ be a {\em positive-definite kernel} or {\em covariance}, i.e., for any ${n\in\Nbb}$, ${a_1,\ldots,a_n \in \Rbb}$ and 
${x_1,\ldots,x_n \in \Xcal}$, we have 
$ \sum_{i=1}^n\sum_{j=1}^n a_i a_j\bbk(x_i,x_j) \ge 0$. 
Accordingly, we say $f$ is a GP with mean function $\mu$ and kernel $k$, denoted by ${f\sim \GP{\mu}{\bbk}}$, if for any finite set of points ${\{x_1,\ldots,x_n\} \subseteq \Xcal}$, the vector ${\bma{@{}c@{\,}c@{\,}c@{}}{f(x_1), & \ldots, & f(x_n)}^\tr}$ has a normal distribution with mean vector ${\bma{@{}c@{\,}c@{\,}c@{}}{\mu(x_1), & \ldots, & \mu(x_n)}^\tr}$ and covariance matrix $K =(K_{i,j})\in$ $\Rbb^{n\times n}$ defined as $K_{i,j} = \bbk(x_i,x_j)$, for all ${i,j\in \Nbb_n}$.

GPs provide prior distributions over the function space $\Rbb^\Xcal$ as well as flexible  classes  of  models suitable for Bayesian learning. 
The marginal and conditional means and variances can be computed in a closed form due to the properties of the Gaussian distributions. 
Accordingly, one can develop a probabilistic  non-parametric regression  method \citep{rasmussen2006gaussian} as briefly discussed in the following.

\subsection{Kriging or Gaussian Process Regression}\label{ssec:GPR}
Let ${f\sim \GP{\mu}{\bbk}}$ and ${\Xcal_{n_\Dcal}:=\{x_1,\ldots,x_{n_\Dcal}\} \subseteq \Xcal}$. Also, for ${\ki=1,\ldots,{n_\Dcal}}$, let  ${y_\ki = f(x_\ki)+w_\ki}$, where ${w_1,\cdots,w_{n_\Dcal}}$ are measurement noise which are independent random variables with distribution ${\Ncal(0,\sigma_w^2)}$.  Define the set of data, $\Dcal$, as ${\{(x_\ki,f(x_\ki))\,|\, \ki=1,\ldots,n_\Dcal\}}$
and vector ${\vc{y}_{n_\Dcal}}$ as ${[y_1,\,...\,,y_{n_\Dcal}]^\tr}$.
Then, for any point ${x_*\in\Xcal}$, we have
\begin{equation}\label{eq:fdist}
f(x_*)|\Dcal,x_* \ \sim \ \Ncal(m(x_*),\sigma^2(x_*)),
\end{equation}
where
\begin{align}
&m(x_*) = \mu(x_*) + \vc{k}(x_*)^\tr (\mx{K}_{n_\Dcal}+\sigma^2_w\eye_{n_\Dcal})^{-1}\vc{y}_{n_\Dcal},\label{eqn:mx_*}\\
&\sigma^2(x_*) = \bbk(x_*,x_*) -\vc{k}(x_*)^\tr (\mx{K}_{n_\Dcal}+\sigma^2_w\eye_{n_\Dcal})^{-1}\vc{k}(x_*),\label{eqn:sigmax_*}  
\end{align}
given that $\vc{k}$ and $\vc{K}_{n_\Dcal}$ are defined as
\begin{equation}\label{eqn:k_star_K_n}
\begin{split}
\vc{k}(x_*)\! =\!\! \begin{bmatrix}\bbk(x_*,x_1)\\\vdots\\\bbk(x_*,x_{n_\Dcal})\end{bmatrix}\!,  
\mx{K}_{n_\Dcal}\!=\!\! \begin{bmatrix}\bbk(x_1,x_1)\!&\!\cdots\!&\!\bbk(x_1,x_{n_\Dcal})\\\vdots\!&\!\ddots\!&\!\vdots\!\\\bbk(x_{n_\Dcal},x_1)\!&\!\cdots\!&\!\bbk(x_{n_\Dcal},x_{n_\Dcal})\end{bmatrix}\!.
\end{split}
\end{equation}

\subsection{Square Exponential Kernel}\label{ssec:SE_kernel}
In \citep{rasmussen2006gaussian}, various candidates were introduced for the choice of kernel. 
If the index set, $\Xcal$, is a subset of $\Rbb^d$, the most common kernel is the {\em square exponential} defined as
\begin{equation}\label{eqn:SEkernel}
\bbk(\vc{x},\vc{x}') = \sigma_f^2 \exp(-\frac{1}{2}(\vc{x}-\vc{x}')^\tr\Lambda(\vc{x}-\vc{x}')), \quad \forall \vc{x},\vc{x}'\in \Xcal, 
\end{equation}
with ${\Lambda = \diag(\lambda_1^2,\cdots,\lambda_d^2)}$ where
${\lambda_1,\cdots,\lambda_d}$ are the length-scale parameters determining the flatness of $f$, and $\sigma_f^2$ is the variance of $f$. These parameters are the {\em hyperparameters} of the kernel. The vector of hyperparameters is denoted by $\theta$ and defined as ${(\lambda_1,\ldots,\lambda_d,\sigma_f)}$.

\subsection{Hyperparameter Estimation}\label{ssec:HyperEst}
A common way to estimate the hyperparameters of the model is maximizing the {\em marginal likelihood}, also known as {\em evidence}. 
More precisely, the hyperparameters are estimated based on the following optimization problem
\begin{equation}\label{eqn:max_marginal_likelihood}
\hat{\theta}:= 
\argmaxOp_{\theta\in\Theta}\ p(\vc{y}_{n_\Dcal}|\Xcal_{n_\Dcal},\theta),
\end{equation}
where $\Theta$ is the feasible set of hyperparameters, and $p$ denotes the probability density function. 
This is equivalent to minimizing the {\em negative log marginal likelihood} (nlml).
Although, this problem is not convex, the derivatives of the objective function can be calculated analytically, see \citep[Chapter 5]{rasmussen2006gaussian}.

\section{Cable Net Model and Problem Formulation}
\label{sec:model}

In this section, we present the model of the cable net system and the problem formulation. 

\subsection{Model of the Cable Net}

The flexible formwork consists of a supporting rigid frame and a cable net which is prestressed inside the frame. 
All edges of the net are in tension and form a doubly curved geometry. 
We define the graph $\mathcal{G} = (\mathcal{V},\mathcal{E})$ to represent the topology of the cable net. Each node $i$ of the index set $\mathcal{V}$ is equipped with the position coordinates $r_i = \bma{@{}c@{\,\,\,\,}c@{\,\,\,\,}c@{}}{r_{x,i} & r_{y,i} & r_{z,i}}^\top$. The position coordinates of all nodes define the form of the cable net. The edges with index set $\mathcal{E}$ represent the cable segments of the net.

We distinguish between the $n_B$ boundary nodes which are positioned on the rigid frame where the cable net is attached, and the $n_I$ interior nodes which lie in the interior of the net. The same distinction is made for the edges, where the $m_B$ boundary edges are directly connected to the rigid frame. The corresponding variables or parameters are denoted by an index $B$ or $I$, respectively. The nodal position coordinates in the interior of the net are therefore denoted by $r_I \in \mathbb{R}^{3 n_I}$ and the ones at the boundary of the net are denoted by $r_B \in \mathbb{R}^{3 n_B}$. 
The lengths of the $m$ cable net edges are defined as the Euclidean distances between the adjacent nodes. The length of the edge between the nodes $s$ and $t$ is thus defined as 
\begin{equation}
l_{(s,t)} := \|r_s - r_t \|_2. 
\end{equation} 
We denote the parameter of unstressed length of the edge $i$ by $l_{0,i}$. 
The edges in the interior of the net are of fixed unstressed lengths. These parameters are stacked in the vector 
\begin{equation}\label{eq:l0I}
l_{0,I} := [l_{0,1}, ..., l_{0,m_I}]^\top \in \mathbb{R}^{m_I}. 
\end{equation}
The boundary edges of the cable net are attached to the rigid frame via turnbuckles. The latter provide some degrees of freedom to apply control inputs in the form of adjustments of the unstressed edge lengths. The control inputs can be applied in order to change the form of the cable net. The parameters of unstressed lengths of the boundary edges are stacked in the vector $l_{0,B} := [l_{0,m_{I+1}}, ..., l_{0,m}]^\top \in \mathbb{R}^{m_B}$, with ${m = m_I+m_B}$. 
The parameter vector containing the unstressed lengths of all cable net edges is defined as 
\begin{equation}
l_0 := [l_{0,I}^\top, \, l_{0,B}^\top]^\top 
\in \mathbb{R}^{m}.
\end{equation} 
Figure~\ref{fig:Skizze} shows a top view of a cable net system. 
\begin{figure}
	\centering\input{figures/Skizze_Cablenet.tex} \label{fig:Skizze}
	\includegraphics[width=0.65\columnwidth]{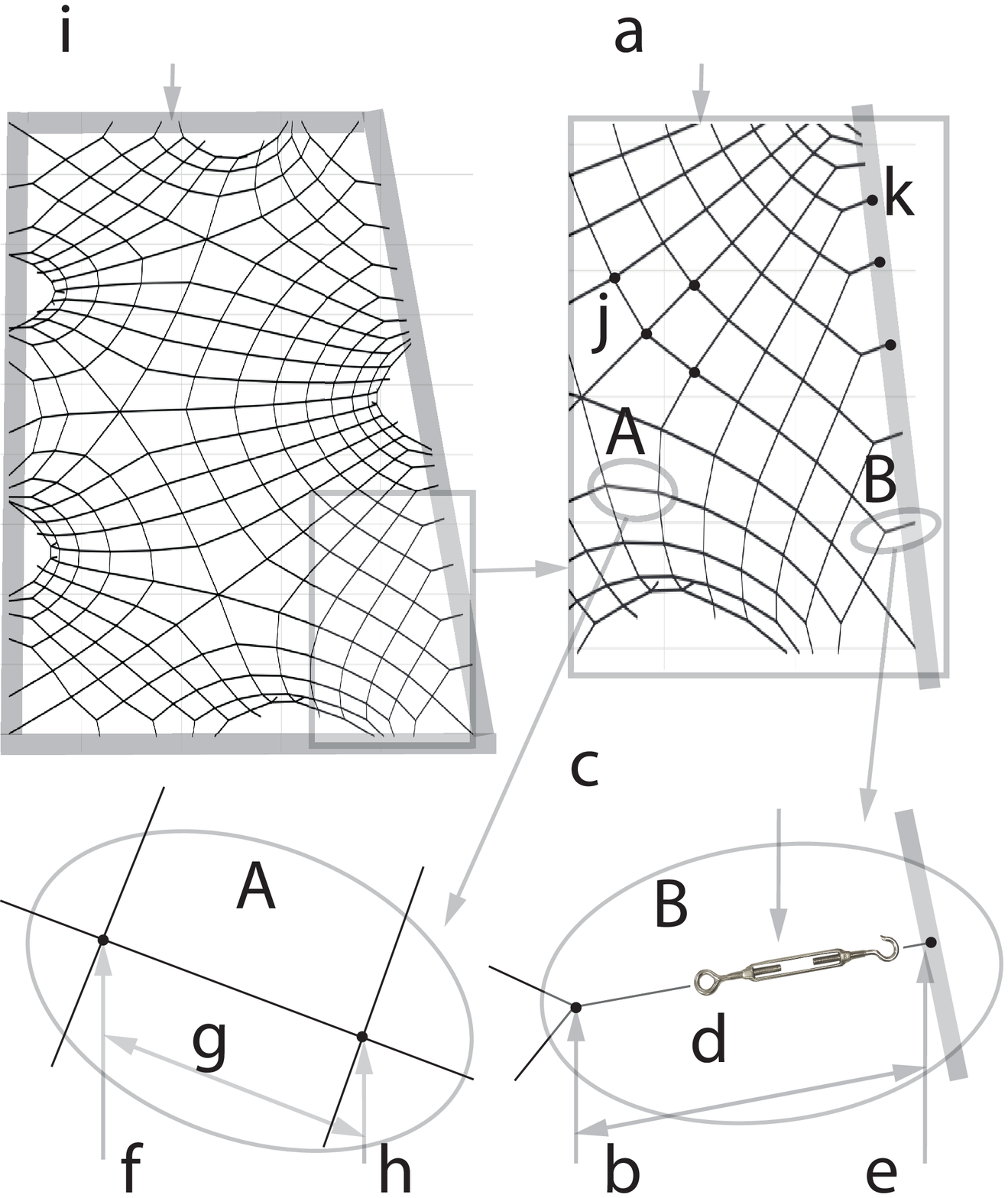}
	\caption{Top view of the cable net system, \citep{Stuerz2019}. a) Interior edge $(s_a,t_a) \in \mathcal{E}_I$ connecting nodes $s_a \in \mathcal{V}_I$ and $t_a \in \mathcal{V}_I$. 
		b) Boundary edge $(s_b,t_b) \in \mathcal{E}_B$ 
		connecting the interior node $s \in \mathcal{V}_I$ and the boundary node $t \in \mathcal{V}_B$. \label{fig:Skizze}}
\end{figure}

\subsection{Force Balance Equations}

The static equilibrium of the cable net can be characterized by the nodal  coordinates $r$ for which the net forces at all interior nodes are zero, i.e., given by the solution of the equations
\begin{equation}
\begin{aligned} \label{eq:hq}
h_{s} 
\!= \!\!\!\!\!\sum_{(s,t) \in \bar{\mathcal{E}}_s} \!\!\!\!\!EA_{(s,t)} \!\left( \begin{bmatrix} r_{x,s} \\ r_{y,s} \\ r_{z,s} \end{bmatrix} \!-\! \begin{bmatrix} r_{x,t} \\ r_{y,t} \\ r_{z,t} \end{bmatrix} \right) \!\left( \frac{1}{{{l}}_{0,(s,t)}} - \frac{1}{l_{(s,t)}}  \right) \!=\! 0, 
\end{aligned}
\end{equation}
for all interior nodes $s \in \mathcal{V}_I$ and for all tensioned edges $(s,t)$ which are adjacent to node $s$, denoted by $\bar{\mathcal{E}}_s$. The material parameter $EA_{(s,t)}$ represents the elasticity of the edge $(s,t)$. 

For fixed boundary values, $r_B$, the implicit function $h: \mathbb{R}^{3 n_I} \times \mathbb{R}^{m_B} \mapsto \mathbb{R}^{3 n_I}$, with ${h(r_I, l_{0,I}) = 0}$, is the vector containing the force equilibrium equations in $x$-, $y$- and $z$-directions for all interior nodes, i.e., 
\begin{equation} 
\label{eq:h}
h(r_I, l_{0,I}) = \begin{bmatrix} h_1^\top & \hdots & h_{n_I}^\top \end{bmatrix}^\top = 0. 
\end{equation} 

For more details of the model, we refer to \citep{Stuerz2019}.

\subsection{Problem Formulation}

Prior work has identified the unstressed edge lengths $l_0$ as the parameters with the largest uncertainties and biggest influence on the cable net form. As the cable net is constructed in a tensioned state, these lengths cannot accurately be measured. 
We make the following assumptions about the system. 
The boundary positions, $r_B$, at the rigid frame are known and constant, i.e., the frame does not flex or bend for changing edge lengths parameters. 
Since the unstressed lengths of the boundary edges, $l_{0,B}$, are actuated, their  desired values 
can be precisely configured. Therefore, their values are known and constant. 
In contrast, the nominal (desired) values of the unstressed lengths of the interior edges,  $l_{0,I}$, 
are subject to fabrication variations and are non-adjustable and therefore cannot be precisely achieved on the as-built system. Furthermore, the parameters $l_{0,I}$ also cannot be directly measured on the system and are therefore uncertain. 
They however have a big influence on 
the form of the cable net and thus on the performance of model-based control methods, and therefore need to be precisely identified. 
In order to reduce the time and cost on the construction site, the number of measurements needed for the identification should be minimized. 

Given the constant boundary position coordinates, $r_B$, and the constant boundary edge lengths, $l_{0,B}$, and given the nonlinear implicit equations characterizing the static equilibrium form of the cable net from \eqref{eq:hq}, 
${h_s(l_{0,I}, r_{I})=0, \forall s \in \mathcal{V}_I}$, 
the function 
\begin{equation} \label{eqn:f}
\begin{array}{c}
\vc{f}: \mathbb{R}^{3 n_I} \to \mathbb{R}^{m_I}, \\
 r_I \mapsto l_{0,I}, 
\end{array}
\end{equation} 
mapping the position coordinates of the interior nodes, $r_I$, into the parameters $l_{0,I}$, is to be found. 
Then, one set of measurements of $r_I$ is sufficient to identify the parameters $l_{0,I}$. 
As the function $\vc{f}$ is not analytically known, we use nonlinear regression to learn it. This procedure is discussed in the sequel.

\section{Gaussian Processes for Cable Net Parameter Identification}
\label{sec:GPlearning}

In this section, we present the learning procedure for the latent function $\vc{f}$.
Using the learned function, denoted here by $\hat{\vc{f}}$, and a single set of measurements of $r_I$, denoted by $\hat{r}_I$, we identify the parameters $l_{0,I}$ of the cable net by evaluating $\hat{l}_{0,I} = \hat{\vc{f}}(r_I)$.

\subsection{Learning Procedure}
\label{ssec:Learning} 

According to \eqref{eqn:f}, for each interior edge $i\in \mathcal{E}_I$, one can define the function $f_i$ as the 
$i^{\text{\tiny{th}}}$ 
coordinate of $\vc{f}$. 
More precisely, we can define functions
\begin{equation}\label{eq:fi}
f_i:\mathbb{R}^{3 n_I} \to \Rbb, \quad \forall i\in \mathcal{E}_I, 
\end{equation}
such that
\begin{equation}
\vc{f}(r_I) := [f_1(r_I),\cdots,f_{m_I}(r_I)]^\tr = l_{0,I}.
\end{equation}

\begin{remark} 
One can show that the function f exists under certain conditions. 
\end{remark}

The latent function, $\vc{f}$, can be learned by learning each of its coordinates, $f_i, i=1,\ldots,m_I$, using an appropriate nonlinear regression method. 
The approach used here is the Gaussian process regression introduced in Section \ref{ssec:GPR}. 
Toward this, we need a suitable kernel or covariance function as well as a set of training data. 
Regarding the kernel, we employ square exponential kernels introduced in \eqref{eqn:SEkernel}.
Note that here $d$ equals $3n_I$. 
In particular, we choose {\em isotropic} square exponential kernels, i.e., we assume that ${\Lambda = \lambda^2 \eye_d}$ and subsequently, for any pair of $x_1$ and $x_2$, we have \begin{equation}\label{eqn:SEkernel_iso}
\bbk(x_1,x_2) = \sigma_f^2 \exp({-\frac{\lambda^2}{2}\|x_1-x_2\|^2}). 
\end{equation} 
This choice is validated by the numerical experiments in Section~\ref{sec:numerics}. 
Accordingly, the hyperparameters of the model are $\lambda$, $\sigma_f$. 
In order to estimate the hyperparameters, we use likelihood maximization or equivalently negative log-likelihood minimization which are non-convex optimization problems (see Section~\ref{ssec:HyperEst}). 
The kernel being isotropic simplifies the optimization problem and also helps to avoid spurious local minima during the negative log-likelihood minimization. 
In addition to kernel selection and hyperparameter estimation, we also need suitable training data. More precisely, we need a set of data denoted by $\Dcal = \{(r_I(\ki),l_{0,I}(\ki)) \,|\, \ki = 1,...,n_\Dcal\}$ which comprises $n_\Dcal$ pairs of $(r_I,l_{0,I})$. 
Let us assume for the moment that we have access to such a data set $\Dcal$. Then, Algorithm~\ref{alg:inv-map} summarizes the steps of the learning procedure described in this section. 
%__________________________________________________
\begin{algorithm}[t]
\caption{GP Regression for Learning the Function $\vc{f}$}\label{alg:inv-map}
\begin{algorithmic}[1]
\State \textbf{Input:} 
Choice of kernel (here: SE in \eqref{eqn:SEkernel_iso})
\State Training data set $\Dcal = \{(r_I(\ki),l_{0,I}(\ki)) \,|\, \ki = 1,...,n_\Dcal\}$
\State \textbf{Initialization:} 
Set mean function to zero: $\mu = 0$ 
\State Take $\vc{k}$ as the SE kernel
\For{$\ki=1,\ldots,n_{\Dcal}$} 
        \State Set $x_\ki := r_I(\ki)$
        \EndFor\label{loop0}
\State \textbf{end}
\State \textbf{Routine:} 
\For{$i = 1,...,m_I$ }
\For{$\ki=1,\ldots,n_{\Dcal}$} 
\State Set $y_\ki :=l_{0,i}(\ki)$ 
\EndFor\label{loop1}
\State \textbf{end}
\State Estimate hyperparameters $\theta^{(i)}:=(\lambda^{(i)},\sigma_f^{(i)})$ as in \eqref{eqn:max_marginal_likelihood}%{eqn:min_nlml} 
\State Compute $\mx{K}_{n_{\Dcal}}$  as in \eqref{eqn:k_star_K_n} and derive function $\vc{k}(\cdot)$ 
\State Derive functions $m^{(i)}(\cdot)$ and $\sigma^{(i)}(\cdot)$ as in \eqref{eqn:mx_*}, \eqref{eqn:sigmax_*}
\EndFor\label{loop2}
\State \textbf{end}
\State \textbf{Output:} 
${\hat{\vc{f}}(\cdot):=[m^{(i)}(\cdot)]_{i\in \mathcal{E}_I}}$ and ${\Sigma(\cdot):=[\sigma^{(i)}(\cdot)]_{i\in \mathcal{E}_I}}$ 
\end{algorithmic}
\end{algorithm}
%__________________________________________________ 
The procedure for obtaining suitable training data 
is presented in the following.

\subsection{Efficient Data Generation} 
\label{ssec:finverse}

In order to train the GP, data pairs $(r_I,l_{0,I})$ are required. 
As measured data is very expensive, simulated data can be used instead. 
The training data can easily be computed making use of the following observation.  
The inverse function $\vc{f}^{-1}$ that maps the parameters $l_{0,I}$ into the nodal coordinates $r_I$, 
\begin{equation} \label{eq:finv}
\begin{array}{c}
\vc{f}^{-1}: \mathbb{R}^{m_I} \to \mathbb{R}^{3 n_I}, \\ l_{0,I} \mapsto r_I, 
\end{array}
\end{equation} 
can be reformulated as a second-order cone program and thus be evaluated by convex programming, as presented in \citep{Stuerz2016a}, \citep{Stuerz2019}. This is true for fixed parameters $l_{0,I}$, $r_B$ and $l_{0,B}$, and for all edges being in tension. 
Under the assumption that the frame does not flex or bend, the position coordinates of the boundary nodes, $r_B$, can be assumed to be constant. They can be measured precisely on the construction site, and are therefore known \citep{Stuerz2019}, \citep{Liew2018}. 
As discussed before, the unstressed boundary edge length parameters, $l_{0,B}$, can precisely be actuated to the desired value and are therefore also known. The parameters $l_{0,I}$ will be randomly chosen for the data generation. 
The underlying second-order cone program that needs to be solved is briefly presented in the following.

In order to find the static equilibrium of the system in terms of the coordinates $r_I$, for given parameters $l_{0,I}$, the approach of minimizing the total energy of the cable net can be taken. We assume that the elastic tension forces versus elongation function of the edges are linear and increasing. Then, for fixed parameters $l_{0,I}$, $l_{0,B}$ and $r_{B}$ the problem of minimizing the total energy of the system is equivalent to the following convex second-order cone program (SOCP) \citep{Stuerz2016a} 
\begin{align} \label{SOCP}
	& &&\min_{{r}_I, v, \w} ~  
	\frac{1}{2} {v}  & && \notag \\
	& && ~ \mathrm{s.t.}  ~~~ \bigg{[}\frac{{EA_{(s,t)}}}{{ {{l}}_{0,(s,t)} }}\bigg{]}^{\frac{1}{2}} ~  \left( \left\Vert  {r}_s  -  {r}_t  \right\Vert_2 - {l}_{0,(s,t)} \right)  \leq \w_{(s,t)} , 
	& && \notag \\
	& && ~~~~~~~~~ 0 \leq \w_{(s,t)}, \quad \forall ~(s,t) \in \mathcal{E}, & && \\ 
	& && ~~~~~~~~\left\Vert \bma{@{}c@{}}{ 2 \w \\ 1- {v} } \right\Vert_2 \leq 1+ {v}\,, & && \notag 
\end{align}
where the vector $\w \in \mathbb{R}^m$ and the variable $v \in \mathbb{R}$ are introduced in order to formulate the problem as an SOCP. 
For details about this problem formulation we refer to \cite{Stuerz2019}.

In order to generate the data set 
$\Dcal = \{(r_I(\ki), l_{0,I}(\ki)) \,|\, \ki = 1,...,n_\Dcal\}$, 
a set of $n_\Dcal$ parameter vectors of unstressed interior edge lengths, $l_{0,I}$, are randomly generated as 
\begin{equation} \label{eq:data}
l_{0_I}(\ki) = \bar{l}_{0,I} + \Delta l_{0,I}(\ki), \quad \text{for~} k = 1, ... , n_\Dcal. 
\end{equation} 
In \eqref{eq:data}, $\bar{l}_{0,I}$ is the vector of nominal parameters designed to achieve the desired geometry of the cable net, 
and ${\Delta l_{0,I}(\ki)}$ are the realized uncertainty vectors, e.g., drawn from a uniform distribution. 
In order to compute the corresponding position coordinates of the interior nodes, $r_I(\ki)$, representing a static equilibrium of the cable net for the given parameters $l_{0,I}(\ki)$, the inverse function $\vc{f}^{-1}(l_{0,I}(\ki))$ as in \eqref{eq:finv} can efficiently be evaluated by solving the SOCP in \eqref{SOCP}. 
The steps of the data generation are summarized in Algorithm~\ref{alg:datagen}. 
%__________________________________________________
\begin{algorithm}[t]
\caption{Generation of Data Set $\Dcal$ 
}\label{alg:datagen}
\begin{algorithmic}[1]
\State \textbf{Input:} 
Size of data set $n_\Dcal$, nominal parameters $\bar{l}_{0,I}$, probability distribution to draw uncertain parameters from e.g.,  $\mathcal{U}(a,b)$, parameters $l_{0,B}$, $r_B$, $EA$, incidence matrix of $\mathcal{G}$ 
\State \textbf{Routine:} 
\For{$\ki=1,\ldots,n_{\mathcal{D}}$} 
        \State Generate random uncertainty $\Delta l_{0,I}$ drawn from $\mathcal{U}(a,b)$ 
        \State \smash{Generate realized uncertain parameter vector} ${l_{0,I} = \bar{l}_{0,I} + \Delta l_{0,I}}$
\State Compute $r_I$ by solving the SOCP in \eqref{SOCP}
\State Set $r_I(\ki) := r_I$ and $l_{0,I}(\ki) := l_{0,I}$
\EndFor\label{loop2}
\State \textbf{end}
\State \textbf{Output:} 
Training data set $\Dcal = \{(r_I(\ki),l_{0,I}(\ki)) | \ki = 1,...,n_\Dcal\}$
\end{algorithmic}
\end{algorithm}
%__________________________________________________

\subsection{Parameter Identification Based on Trained GP}
\label{ssec:ident}
After generating the data set $\Dcal$ as in Algorithm~\ref{alg:datagen} and training the GP to learn the function $\hat{\vc{f}}$ as in Algorithm~\ref{alg:inv-map}, we are ready for the parameter identification of the unstressed interior edge lengths $l_{0,I}$. 
On the construction site, the values of $r_I$ can be measured in a very accurate way \citep{Stuerz2019}, \citep{Liew2018}. 
Based on one measurement of the interior nodal coordinates of the cable net, which we denote by $r_I^{\mathrm{meas}}$ in the following, the parameter identification of $l_{0,I}$ is 
then given by 
\begin{equation} \label{eq:ident}
\hat{l}_{0,I} := \hat{\vc{f}}(r_I^{\mathrm{meas}}). 
\end{equation}

\section{Numerical Experiments}
\label{sec:numerics}

\begin{figure}[t] 
	\centering
	\input{figures/3Dplot_frag.tex} 
	\includegraphics[width=0.7\columnwidth]{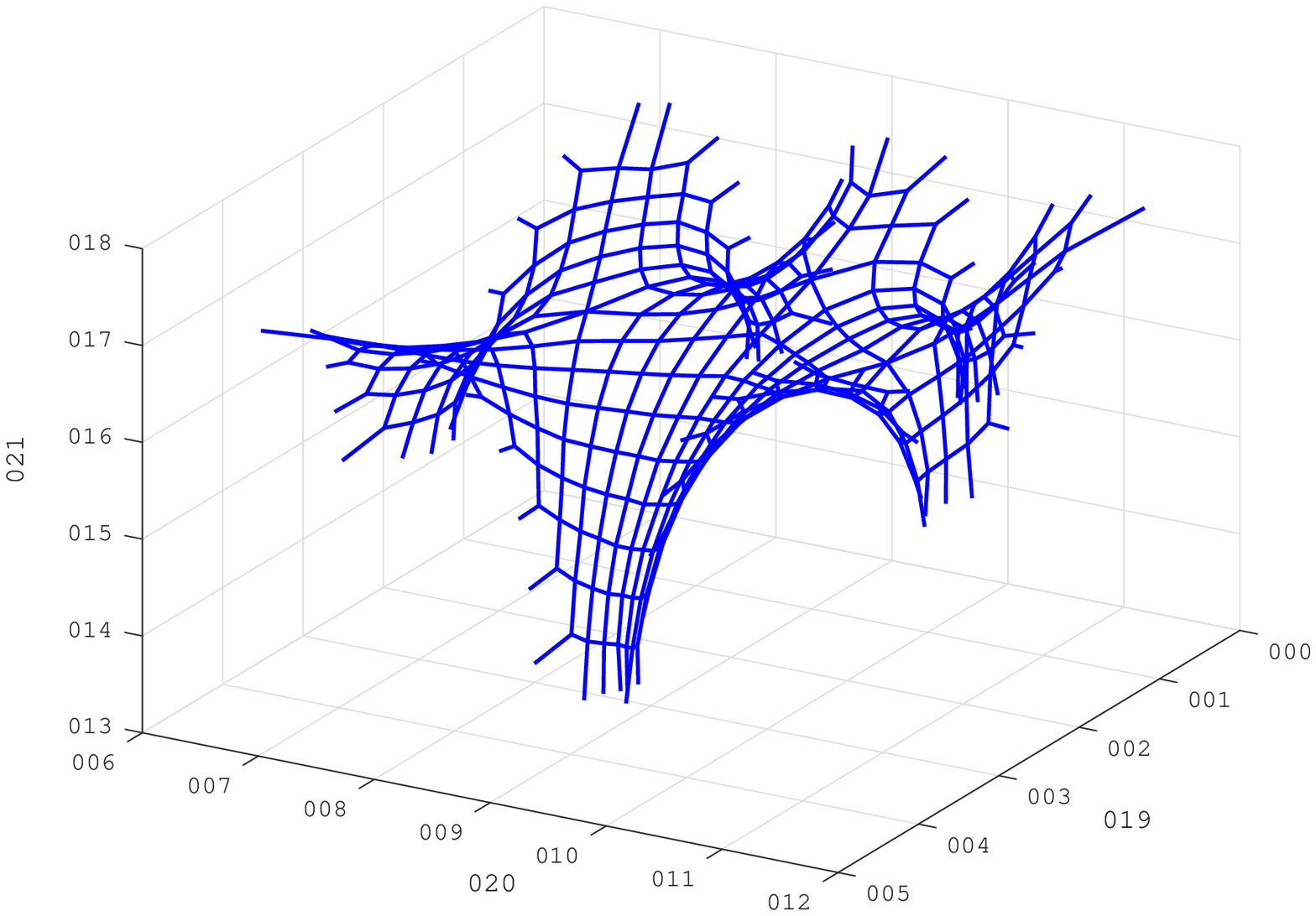} 
	\vspace{-0.1cm}
	\caption{\label{fig:3D} Cable net system inspired by  
	\citep{Block2017, Stuerz2019}.}
\end{figure}

We present numerical results on a cable net formwork for the construction of a doubly curved lightweight roof structure, inspired by a quarter-scale prototype of the HiLo roof 
on the NEST building at the EMPA campus in D\"ubendorf, Switzerland \citep{Block2017}. % 
The prototype model has ${m_I = 536}$ interior edges, and thus $l_{0,I} \in \mathbb{R}^{536}$. 
The number of interior nodes is ${n_I = 300}$, and therefore $r_I \in \mathbb{R}^{900}$. 
More details about the prototype are given in \citep{Stuerz2019}.  
Figure~\ref{fig:3D} shows a plot of the cable net. 
The nominal values of the unstressed edge lengths $l_{0,B}$ and $\bar{l}_{0,I}$ are known. They result from a design problem to match the desired target form of the cable net. While the boundary edge lengths $l_{0,B}$ are adjustable, and can therefore be precisely set during the installation of the cable net system, the interior unstressed edge lengths $l_{0,I}$ are not adjustable. We assume that the true parameters $l_{0,I}$ lie within ${\pm 5\,\mathrm{mm}}$ of their nominal values $\bar{l}_{0,I}$. 

In the following, instead of identifying the whole parameter vector $l_{0,I}$ as presented before, we identify the uncertainty $\Delta l_{0,I}$. 
In addition to the nominal parameters,  $\bar{l}_{0,I}$, and the true parameters ${l}_{0,I}$, as in \eqref{eq:data}, we further define the predicted parameters, $\hat{l}_{0,I}$, as
\begin{equation}\label{eq:l0truenompred} 
\begin{aligned}
    \hat{l}_{0,I} := \bar{l}_{0,I} + \Delta \hat{l}_{0,I}. 
\end{aligned}
\end{equation}
For fixed parameters $r_B$ and $l_{0,B}$, let us define the nominal form $\bar{r}_I$, the true form ${r}_I$, and the predicted form $\hat{r}_I$, which are the solutions to the SOCP in \eqref{SOCP} for the corresponding parameters $\bar{l}_{0,I}$, ${l}_{0,I}$, and $\hat{l}_{0,I}$, respectively. 
We further define the following deviations from the nominal form 
\begin{equation}
\begin{aligned}
    \Delta {r}_I := r_I - \bar{r}_I, \qquad 
    \Delta \hat{r}_I := \hat{r}_I - \bar{r}_I. 
\end{aligned}
\end{equation}
With these definitions, the function that is learned in Algorithm~\ref{alg:inv-map} is ${\vc{f}: \Delta r_I \mapsto \Delta l_{0,I}}$. 

A set of $n_{\Tcal} = 1000$ data pairs $(\Delta r_I, \Delta l_{0,I})$ is generated using Algorithm~\ref{alg:datagen} in Section~\ref{ssec:finverse}. 
The realized uncertainty vectors are randomly generated, drawn from a uniform distribution of ${\pm 5\,\mathrm{mm}}$, i.e., ${\Delta l_{0,I}(\ki) \in \mathcal{U}(-0.005,0.005)}$. 
We divide the data set into ${n_{\Dcal} = 950}$ data pairs that will be used for the training of the GP, and ${n_{\Vcal} = 50}$ data pairs that will be used for cross-validation. 
The GP is then trained as presented in Algorithm~\ref{alg:inv-map} in Section~\ref{ssec:Learning} to learn the function ${\vc{f}}:\Delta r_I \mapsto \Delta l_{0,I}$. 

Each of the $n_{\Vcal}$ points $\Delta r_I(\ki)$, $\ki = 1,...,n_{\Vcal}$ from the validation data set is then used to represent a measurement of the deviation between the true cable net form and the nominal one. 
For each of the points $\Delta r_I(\ki)$ 
the parameter vector ${\Delta \hat{l}_{0,I}(\ki)}$ 
is identified via equation~\eqref{eq:ident}. 
The predicted values ${\Delta \hat{l}_{0,I}(\ki)}$ are then compared to the true realized deviations $\Delta {l}_{0,I}(\ki)$ from the validation data set. 
The prediction error for the uncertainty of the unstressed length of edge $i$ of data point $\ki$ is defined as 
\begin{equation}\label{eq:ei}
e_i(\ki) := \Delta l_{0,i}(\ki) - \Delta \hat{l}_{0,i}(\ki), \quad i \in \mathcal{E}_i, \quad \ki = 1,...,n_\Vcal. 
\end{equation} 

Figure~\ref{fig:errorplot_q1} shows the values of $\Delta l_{0,i}(\ki)$, $\Delta \hat{l}_{0,i}(\ki)$ and $e_i(\ki)$ over all interior edges ${i=1,...,m_I=536}$ 
for one randomly chosen data pair $(\Delta r_I(\ki), \Delta l_{0,I}(\ki))$, with $\ki$ randomly fixed to ${\ki=1}$. 

\definecolor{mycolor1}{rgb}{0,0.75,0.75}%
\definecolor{mycolor2}{rgb}{0.75,0.75,0}%
\begin{figure}
	\centering
	\input{figures/errorplot_Y_I_L0_I_0_1_q2.tex} 
	\includegraphics[width=0.68\columnwidth]{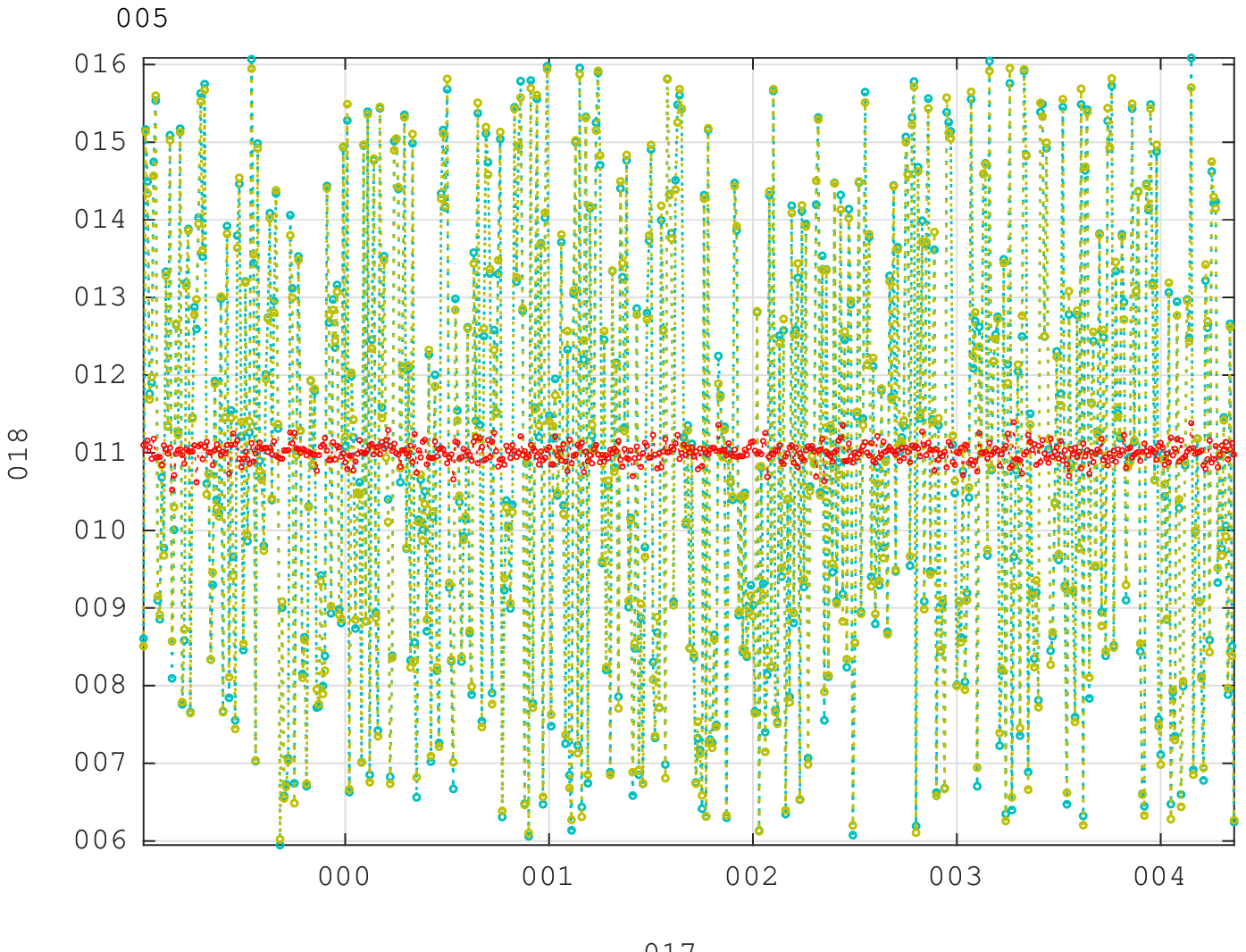} 
	\vspace{0.2cm}
	\caption{ \label{fig:errorplot_q1} 
True and predicted parameter uncertainties for one validation data point, i.e., ${\Delta l_{0,i}(\ki)}$ and ${\Delta \hat{l}_{0,i}(\ki) = \hat{\vc{f}}(\Delta l_{0,i}(\ki))}$ with randomly chosen $\ki=1$ over all interior edges ${i=1,...,m_I}$: ~~
\textcolor{mycolor1}{\textbf{$\cdot \cdot \bullet \cdot \cdot$}} ~ Predicted $\hat{\Delta l_{0,i}}$ ~~ 
\textcolor{mycolor2}{\textbf{$\cdot \cdot \bullet \cdot \cdot$}} ~ True ${\Delta l_{0,i}}$ ~~ 
\textcolor{red}{\textbf{$\cdot \cdot \bullet \cdot \cdot$}} ~ 
Error $e_i$ as in \eqref{eq:ei} for ${\ki=1}$ }
\end{figure} 

In Figure~\ref{fig:boxplot}, we present statistical results for the prediction error $e_i$ over all interior edges $i=1,...,m_I$ considering all ${\ki = 1,...n_\Vcal = 50}$ validation data points. 
The mean of the prediction error $e_i$ over all the $n_\Vcal$ data points of the validation data set is given by 
\begin{equation}\label{eq:m_ei}
    \mathrm{mean}(e_i) := \frac{1}{n_\Vcal} \sum_{\ki=1}^{n_\Vcal}  e_i(\ki), \quad i \in \mathcal{E}_i. 
\end{equation}
Similarly, the standard deviation of the error is obtained as
\begin{equation}\label{eq:std}
    \mathrm{std}(e_i) := \bigg{(}\!\frac{1}{n_\Vcal}\!\sum_{\ki=1}^{n_\Vcal} \left( e_i(\ki)
    \!-\! \mathrm{mean}(e_i)\right)^2 \bigg{)}^{\frac{1}{2}}\!\!\!, \quad i \in \mathcal{E}_i. 
\end{equation} % 
Note that the data in Figure~\ref{fig:boxplot} is plotted on a scale of the $y$-axis of one order of magnitude smaller than the one in Figure~\ref{fig:errorplot_q1}.

For further statistical evaluation of the cross-validation, let us define the prediction error vector for all parameter uncertainties of the interior edge lengths as 
\begin{equation}\label{eq:e}
e(\ki) := \Delta l_{0}(\ki) - \Delta \hat{l}_{0}(\ki), \quad \ki = 1,...,n_\Vcal. 
\end{equation} 
Then, the mean squared error (MSE) and the mean relative squared error (MRSE) over the validation data set are %defined as 
\begin{equation}\label{eq:MSE}
\mathrm{MSE}(e) := \frac{1}{m_I} \sum_{i=1}^{m_I} \left( \frac{1}{n_\Vcal} \sum_{\ki=1}^{n_\Vcal} \left( e_i(\ki) \right)^2 \right),
\end{equation} 
\begin{equation}\label{eq:MRSE}
\text{and} \quad \mathrm{MRSE}(e) := \frac{1}{m_I} \sum_{i=1}^{m_I} \left( \frac{ \sum_{\ki=1}^{n_\Vcal} \left( e_i(\ki) \right)^2}{ \sum_{\ki=1}^{n_\Vcal} \Delta {l}_{0,i(\ki)}^2} \right), 
\end{equation} 
respectively. These metrics have been computed as $\mathrm{MSE} = 1.7\times10^{-8} \mathrm{m}^2$ and $\mathrm{MRSE} = 0.002 = 0.2 \%$, respectively.

\definecolor{mycolor3}{rgb}{0,1,1}
\begin{figure}%[t] 
	\centering\input{figures/boxplot2.tex} 
	\caption{\label{fig:boxplot} 
	\textcolor{black}{$\cdot$} ~ Prediction errors $e_i(\ki), ~\ki = 1,...,n_\Vcal$, as in \eqref{eq:ei} \\
	\textcolor{mycolor3}{---} ~ Mean of prediction error $\mathrm{mean}(e_i)$ as in \eqref{eq:m_ei}  \\
	\textcolor{white!50!blue}{---} ~ $\pm\,\mathrm{std}$ Confidence interval as in \eqref{eq:std}
		}
\end{figure}

In the following, we illustrate the importance of the parameter identification in terms of the impact of the parameter uncertainty $\Delta l_{0,I}$ on the accuracy of the cable net model form $r_I$.  
For demonstration purposes, we focus again on one specific data pair from the validation set, and we choose the one that has been used before in Figure~\ref{fig:errorplot_q1}, i.e., ${\ki=1}$.

\begin{figure} 
	\centering
	\begin{scriptsize}
		\input{figures/surf_nom_2.tex}
		\includegraphics[trim = 3mm 0mm 27mm 0mm, clip, height=7.5cm]{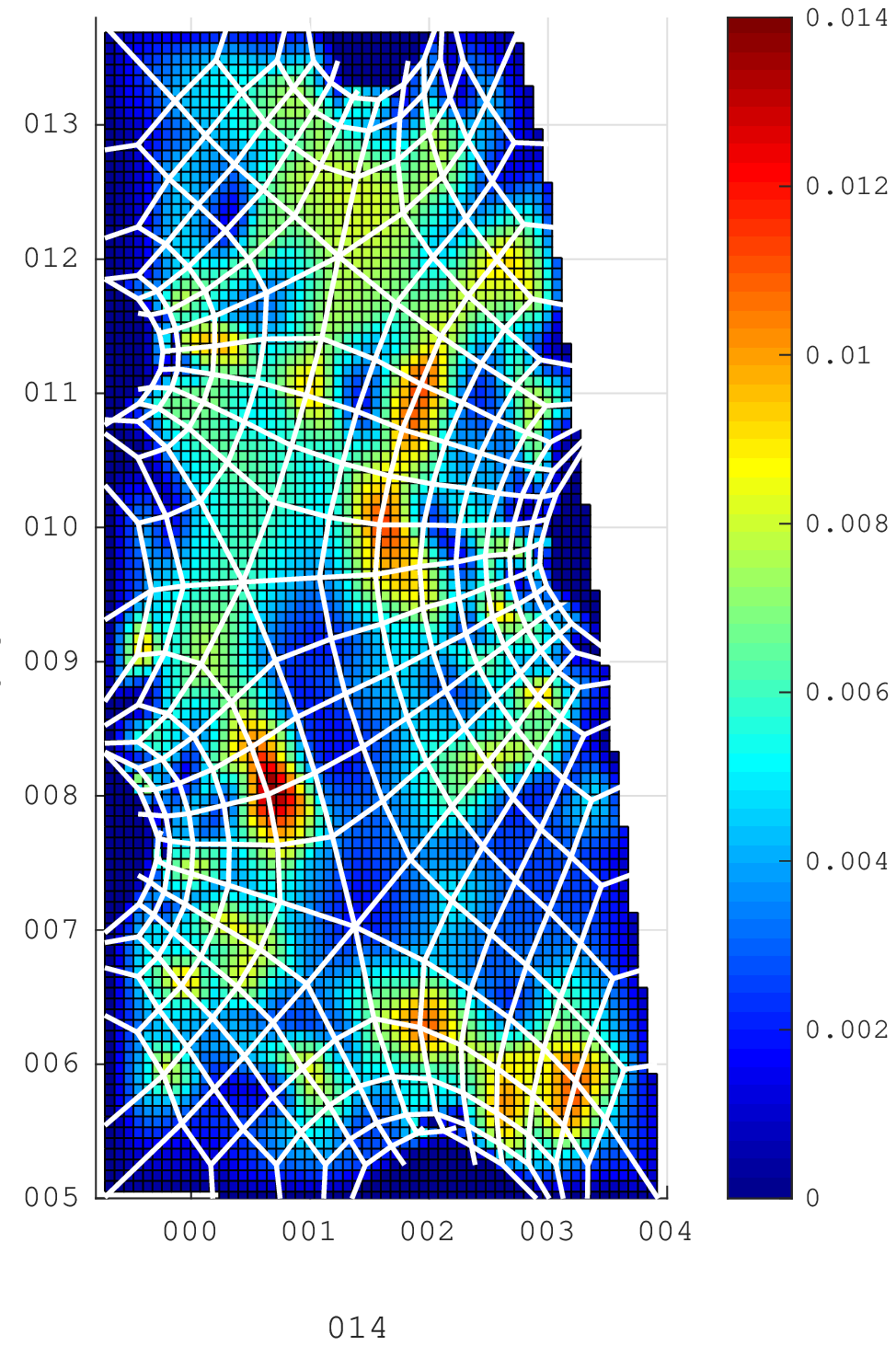}
        %~
		\input{figures/surf_ident_2.tex}
		\includegraphics[trim = 5mm 0mm 0mm 0mm, clip, height=7.5cm]{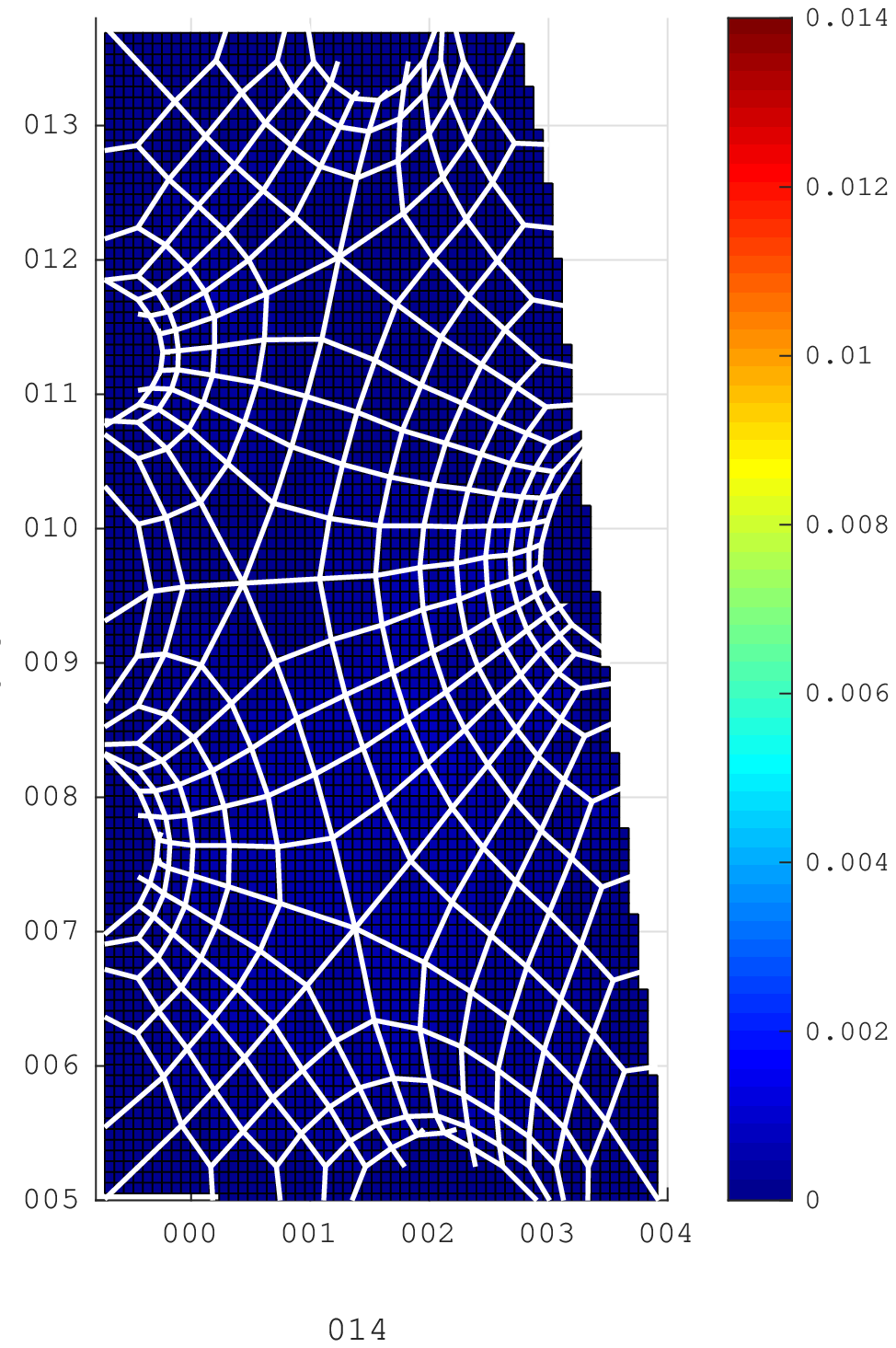}
		\caption{
		 Deviations [m] $\enom$ (left) and $\eident$ (right) as in \eqref{eq:enomident}, between the forms resulting from the nominal and true parameters, $\bar{l}_{0,I}$ and ${l}_{0,I}$, (left), and between the predicted and true parameters, $\hat{l}_{0,I}$ and ${l}_{0,I}$, (right). \label{fig:deviations}}
	\end{scriptsize}
\end{figure}

We compare the deviations between the forms $\bar{r}_I$ and $r_I$ (nominal, i.e., from the model without identification, and true), and between $\hat{r}_I$ and ${r}_I$ (predicted, i.e., from the model with identification, and true). We define the errors $\enom$ and $\eident$ as the vectors of Euclidean distances between the respective position coordinates of the interior nodes ${i = 1,...,n_I}$, i.e., for each $i$ the entries of the vectors $\enom$ and $\eident$ are defined as 
\begin{equation}  \label{eq:enomident}
\begin{aligned}
\enom_i := \|r_i - \bar{r}_i\|_2, \qquad 
\eident_i := \|r_i - \hat{r}_i\|_2. 
\end{aligned} 
\end{equation}
Table~\ref{tab:errors} gives some characteristic distances between these forms, where $\mathrm{RMSE}$ stands for the root mean squared error. 
\captionsetup{width=10cm}
\begin{table}%[h!]
				\captionof{table}{Error Statistics on $\enom$ and $\eident$}
		\begin{center}
		\begin{tabular}{  c  c  c  c  c   }
			\hline
			 & $\|e\|_\infty$ & $\underset{i}{\mathrm{min}}$($|e_i|$) & $\mathrm{mean}(e)$ & $\mathrm{RMSE}$  \\ \hline
			$\enom[\times 10^{-4}]$ & $138$ & $1.93$ & $47$ & $53$  \\ 
			$\eident[\times 10^{-4}]$ & $7.76$ & $0.194$ & $2.73$ & $3.15$\\ \hline
		\end{tabular}
		\label{tab:errors}
	\end{center}
\end{table}
Figure~\ref{fig:deviations} shows these distances over the cable net, i.e., the corresponding entries of the vectors $\enom$ and $\eident$ are shown over the respective nodes. 
This shows the influence of the uncertainty $\Delta {l}_{0,I}$ on the cable net form and therefore the importance of the precise identification of $\Delta \hat{l}_{0,I}$. 
In particular, the latter has a big impact on the performance of model-based control methods for adjusting the cable net form to minimize deviations from the designed target form \citep{Stuerz2019}. 

\section{Conclusion}
\label{sec:conclusion}

This paper presents a novel identification method for important uncertain parameters of a tensioned cable net system, which can be used as a formwork for efficient lightweight construction. 
The parameter identification method is based on GP regression. 
We formulate the function from the form of the cable net to the model parameters and leverage the convexity of the inverse function in order to efficiently generate training data in simulation. 
The impact of the precise parameter identification on the precision of the model-based form of the cable net has been illustrated in numerical experiments.

\bibliography{bib1}

\end{document}